\documentclass[twocolumn,showpacs,preprintnumbers,prb,fleqn,floatfix]{revtex4}

\usepackage{color}
\usepackage{graphicx}
\usepackage{dcolumn}
\usepackage{bm}
\usepackage{amsmath, amssymb}

\begin{document}

\title{Magnetic-field-induced Stoner transition in a dilute quantum Hall system}

\author{B. A. Piot$^{1}$ and D. K. Maude$^{1}$}
\affiliation{$^{1}$ Laboratoire National des Champs Magn\'etiques
Intenses, CNRS-UJF-UPS-INSA, F-38042 Grenoble, France}

\date{\today}

\begin{abstract}
In a recent paper [Phys.Rev.B.\textbf{84}, 161307 (2011)],
experimental data on spin splitting in the integer quantum Hall
effect has been reported in a high mobility dilute 2D electron gas
with electron density as low as 0.2 $\times$ 10$^{11}$ cm $^{-2}$.
In this work, we show that an excellent \emph{quantitative}
description of these data can be obtained within the model of the
magnetic-field-induced Stoner transition in the quantum Hall
regime. This provides a powerful tool to probe the non-trivial
density dependance of electron-electron interactions in the dilute
regime of the 2D electron gas.

\end{abstract}

 \maketitle


The integer quantum Hall effect (IQHE) observed at low temperature
in a 2-dimensional electron gas (2DEG) subjected to a
perpendicular magnetic field can partly be interpreted using a
single particle picture. The Landau quantization gives rise to
cyclotron gaps in the electronic density of states, and the
longitudinale resistance goes to zero each time the number of
occupied Landau level (the filling factor $\nu$) is an even
integer. The odd, or ``spin-resolved'' integer quantum Hall
effect, characterized by the occurrence of zero resistance states
at \emph{odd} filling factors is, however, essentially a
\emph{many-body} phenomenon. Following the seminal theoretical
work of Fogler and Shklovskii,\cite{Fogler95} we have shown that
the lifting of the electron spin degeneracy in the integer quantum
Hall effect at high filling factors should be interpreted as a
magnetic-field-induced Stoner transition.\cite{Piot05} A simple
model with no free parameters correctly predicts the magnetic
field required to observe spin splitting in the longitudinale
resistance, confirming that the odd IQHE is a result of a
competition between the disorder induced energy cost of flipping
spins and the exchange energy gain associated with the polarized
state. This can be thought of as a Stoner transition, since the
only role played by the magnetic field is to modify the density of
states at the Fermi energy. More generally, the absence of
ferromagnetism in a 2DEG at zero magnetic field is related to the
electron-electron correlations which prevent the Hartree-Fock spin
susceptibility from diverging.

Recently, interesting new experimental data on the emergence of
spin splitting in the integer quantum Hall effect has been
reported \cite{Pan11} in a GaAs 2DEG where the density $n_{s}$ can
be tuned down to a value as low as 0.2 $\times$ 10$^{11}$ cm
$^{-2}$, while preserving a high mobility ($\mu>$2 $\times$
10$^{6}$ cm$^{2}$/V s). This data extends the study of spin
splitting to a lower electron density range ( $n_{s} \ll $
10$^{11}$ cm $^{-2}$), where the role of many-body effects is
enhanced and more complex. The aim of this brief report is to show
that an excellent \emph{quantitative} description of these data
can be obtained within the Stoner model. \cite{Piot05}

To describe the appearance of spin splitting quantitatively, the
Fogler and Shklovskii ``half-resolved spin splitting'' criteria
has been retained consistently with previous work. This
corresponds to the condition $\delta\nu=0.5$, where $\delta\nu$ is
the filling factor difference  between two consecutive resistance
maxima in $R_{xx}(B)$ related to spin up and down sub-levels
associated with a given Landau level. According to the Stoner
model, the critical odd filling factor $\nu_{c}$ (magnetic field
$B_{c}=n_{s}h/e \nu_{c}$) for a half-split Landau level
\emph{i.e.} with $\delta\nu=0.5$ must verify in the zero
temperature limit the following set of self-consistent equations:
\cite{Piot07} for the total spin gap $\Delta_{s}$,

\begin{equation}\label{eqn:spingap}
\\ \Delta_{s}=|g^{*}|\mu_{B}B+ X \frac{1}{2} (eB_{c}/h),
\end{equation}
and for the spin polarization within the Landau level,

\begin{equation}\label{eqn:polarisation}
 \delta\nu = \frac{1}{2} = \frac{1}{eB_{c}/h}  \int_{-\infty}^{0} \left[D\left(E+ \frac{\Delta_{s}}{2}\right) -D\left(E-
 \frac{\Delta_{s}}{2}\right) \right] dE,
\end{equation}
where $X$ is the exchange energy between two spins, essentially
depending only on the electron density, and
$D(x)=(1/\Gamma\sqrt{\pi})\exp (-x^{2}/\Gamma^{2})$ is the
normalized density of states for a Gaussian broadened spin Landau
level of full-width at half-maximum $2\sqrt{ln(2)}\Gamma$.

\begin{figure}[h]
\includegraphics[width=1\linewidth,angle=0,clip]{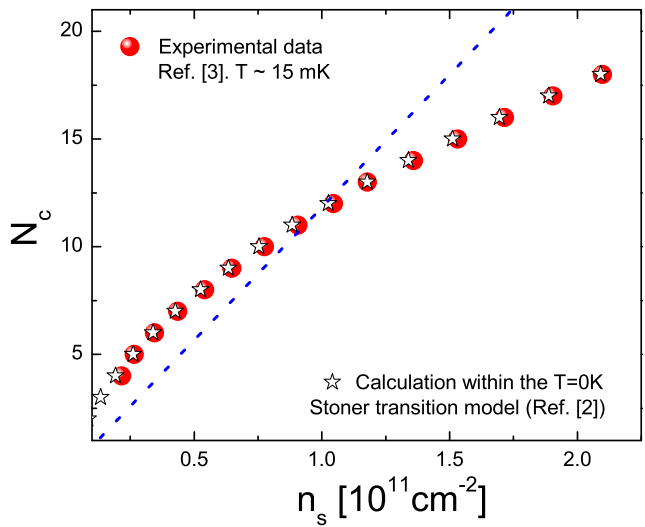}
\caption{(Color online) Critical Landau level index $N_{c}$ at
which the spin splitting is resolved (see text) as a function of
the electron density $n_{s}$. Experimental results of
Ref.~[\onlinecite{Pan11}] (red dots) and calculations within the
Stoner transition model for spin splitting (see text) (open
stars). The dotted line is the result for a fixed (density
independent) value of the exchange parameter $X$.}\label{fig1}
\end{figure}

In figure \ref{fig1}, we plot the critical Landau level index,
$N_{c}=(\nu_{c}-1)/2$, at which the spin splitting is resolved for
a given electron density $n_{s}$. Note that in this plot the
magnetic field is not constant but equal to $n_{s}h/e\nu_{c}$. The
red dots are the experimental results of Ref.~[\onlinecite{Pan11}]
(figure 3). It can be seen that the critical filling factor
$\nu_{c}$ shifts down as the density is reduced. This is because
for a given strength of the exchange interaction and a given
amount of disorder, there is a critical magnetic field which is
required for the spin splitting to be resolved.\cite{Piot05} The
open stars are the results obtained using Eqs. \ref{eqn:spingap}
and \ref{eqn:polarisation}. The density dependant exchange
parameters $X(n_{s})$ has been calculated with the same approach
as in Ref.~[\onlinecite{Piot05}], by using a theoretical spin
susceptibility including finite thickness corrections
\cite{Zhang05} which are non-negligible for the wide 2DEG studied
in Ref.~[\onlinecite{Pan11}]. The only remaining adjustable
parameter, the Landau level broadening $\Gamma$ is found to give
the best fit for  $\Gamma= 0.457$ K ($\sim 0.04$ meV).

An excellent agreement between experiment and theory is obtained,
showing crucially that the Stoner transition describes spin
splitting accurately down to the dilute regime investigated
here.\cite{collapsens} The effect of a non-zero Zeeman energy is
included and shifts the phase transition to higher Landau level
index. The difference between the $g^{*}=-0.44$ and $g^{*}=0$
results (not shown for clarity) is at most $\sim 10$\%, confirming
the secondary role of the Zeeman energy in GaAs in the
perpendicular magnetic field configuration.\cite{Piot07,Leadley98}
Nevertheless, we stress that the Zeeman energy has to be included
to obtain such a good qualitative and quantitative agreement with
experimental data.

The resulting value of the Landau level broadening, $\Gamma=
0.457$ K, is reasonable given the high quality of the 2DEG of
Ref.~[\onlinecite{Pan11}]. We have checked that the density
dependence of the transport lifetime (which can be estimated from
figure 1 in of Ref.~[\onlinecite{Pan11}]) has almost no influence
on our results. The corresponding Landau level ``transport
broadening'' is at most $\Gamma_{tr}= 0.04$ K, which is still an
order of magnitude below the total Landau broadening used here.
This is related to the scattering being essentially long-range in
such high mobility samples. One should mention that, as exchange
interaction and disorder play an opposite role, any error made in
estimating the exchange energy will have an influence on the
resulting value of $\Gamma$. Ideally, these parameters should be
independently determined by complimentary measurements on a given
sample, as detailed in Ref.~[\onlinecite{Piot05}].

In Ref.~[\onlinecite{Pan11}], the experimental data are compared
with the Fogler and Shklovskii (FS) theoretical result obtained
using Eq.4c of Ref.~[\onlinecite{Fogler95}] with the parameters of
the investigated sample. Theory gives a good qualitative
description of the density dependence of $N_{c}$ but lies $\sim
30$\% below the experimental data. There are several possible
reasons for this discrepancy. We recall the main differences
between our model and the theory of FS. Firstly, the Zeeman energy
is not included in Eq.4c of Ref.~[\onlinecite{Fogler95}], which
can lead as we mentioned above to a difference of $\sim 10$\%.
Secondly, the exchange energy in our work is taken from
calculations of the spin susceptibility showing a good agreement
with experimental data for electron densities much lower than
10$^{11}$ cm $^{-2}$ (see figure 3 in
Ref.~[\onlinecite{Zhang05}]). In the theory of FS, it is stressed
that the result is obtained within the approximation $k_{F}a_{B} >
1$, where $k_{F}$ is the Fermi wave factor and $a_{B}$ is the
effective Bohr radius of the 2DEG in GaAs, which corresponds to an
electron density $n_{s} > 1.8 \times$ 10$^{11}$ cm $^{-2}$.
Finally, disorder is taken into account in our approach with an
adjustable parameter ($\Gamma$), whereas the disorder contribution
in Ref.~[\onlinecite{Pan11}] is calculated from Eq.4c of
Ref.~[\onlinecite{Fogler95}] with experimental parameters
extracted for the investigated sample.

We now focus on the behavior of $N_{c}$ in the low density region
($n_{s} < $10$^{11}$ cm $^{-2}$). Interestingly, In this region,
$N_{c}(n_{s})$ exhibits a pronounced sub-linearity. This behavior
is very well caught by the model provided the density dependence
of the exchange interaction is taken into account. To illustrate
this we also plot the result obtained using Eqs. \ref{eqn:spingap}
and \ref{eqn:polarisation} with a fixed value of $X$, taken to be
its value at $n_{s}$= 1 $\times$ 10$^{11}$ cm $^{-2}$ (dotted
line). No concave trend is predicted in this case at low
densities. This can be related extremely simply to the prediction
of the critical magnetic field for the \emph{appearance} of spin
splitting in Ref.~[\onlinecite{Piot05}], corresponding to the case
$\delta\nu=0$:

\begin{equation}\label{eq12}
\\ B_{ss}
= (n_{s}h) /(e \nu_{ss}) =\frac{h\Gamma\sqrt{\pi}}{e}
\frac{1}{X(n_{s})}.
\end{equation}
The critical filling factor $\nu_{ss}$ for the appearance of spin
splitting should be a linear function of the density $n_{s}$ with
a slope $\propto\frac{X}{\Gamma}$, unless the exchange parameters
$X$ or the disorder broadening $\Gamma$ depends on density. The
experimental behavior at low density can thus be attributed to the
theoretically predicted increase in exchange energy as the density
is reduced (see e.g. Ref.~[\onlinecite{Zhang05,DePalo05}]) giving
rise to the concavity of $N_{c}(n_{s})$. The low density behavior
of spin splitting in the integer quantum Hall regime combined with
the Stoner transition approach thus provides a powerful tool to
probe the non-trivial density dependance of electron-electron
interactions in the dilute regime.

To conclude, we have shown that the magnetic-field-induced Stoner
transition approach previously used to describe the spin-resolved
quantum Hall effect can be extended to electron densities as low
as 0.2 $\times$ 10$^{11}$ cm $^{-2}$. In this dilute regime, the
so-called $r_{s}$ parameter corresponding to the ratio of the
Coulomb energy to the kinetic energy reaches $r_{s}\sim 4.3$. In
this situation electron-electron correlations effects become
important and can be experimentally probed by studying the low
magnetic field spin splitting in the quantum Hall regime.

We would like to thank W. Pan and M.M. Fogler for useful
discussions.


\begin{thebibliography}{}
\bibitem[{for()}]{collapsens}
\bibinfo{note}{We additionally note that the collapse of spin splitting as a function
of the electron density within a \emph{given} Landau, as reported
in Refs.~[\onlinecite{Pan11,Wong97}], can be described with the
same success.}
\end{thebibliography}
\end{document}